# Supporting Abstract Relational Space-Time as Fundamental without Doctrinism Against Emergence

Sascha Vongehr

[††]National Laboratory of Solid-State Microstructures, Nanjing University, Hankou Lu 22, Nanjing 210093, P. R. China

Modern physics, via the standard model with Higgs mechanism and string theory for example, has supplied ether-like models and emergent general relativity scenarios that substantially weaken the usual defense of orthodox relativity and abstract, relational space-time in general. Over a dozen arguments in support of relativism against space-substances are discussed. It is not known whether perceived space-time is fundamental or due to a condensed state or string theoretical membrane. Emergent relativity indicates perhaps a whole tower of more fundamental space-times. Whether space is best described as a thing, an emergent phenomenon, or instead Kant's necessary, a priori pre-condition of the possibility of all phenomena, depends on which space-time is under consideration. Nevertheless, space-time is *fundamentally* abstract relational. This is supported by throughout acknowledging as well as illustrating the value of space-substance models, yet still keeping firm in refusing them on the deepest level. The gist is that even while refusing, physics and philosophy gain more from ether concepts than from plainly refusing them.

# 1 Introduction

Mass in general relativity (GR) is simply inertia. Gravity is not a force but curvature of space-time. On the other hand, the Higgs mechanism derives the rest mass of some of the standard model's fundamental particles as an interaction with a 'sticky' background Higgs field through which the particles move. Orthodox relativity almost vilifies such ether-like concepts, while modern physics reintroduced them for about half a century now. In GR, space-time is '*dynamic*', which means that it interacts[a] with the energy-momentum distribution. Nevertheless, orthodox GR can be described as a "relativity is kinematics" position[1,2], i.e. an *abstract-view* (relationism, structuralism) where the dynamics emerges from symmetries (consistency) rather than from any concrete mechanism. The idea that relativistic kinematics arises from the *dynamics* of objects interacting with their background space is disfavored by orthodox GR, because such an interpretation treats the background as too "substantiated"; it seems to resurrect the background as a medium or ether, while background independence is the mark of distinction of GR. High energy particle physics and modern emergent gravity scenarios clash with this orthodox position. String theory even introduced the universe on a membrane and variable light velocities, but the impression of suggesting an ether theory is carefully avoided, because such can still be career suicide. Only physicists who were established beyond reproach could discuss ether-like aspects openly, like George Chapline, Gerd 't Hooft, Robert Laughlin, or Frank Wilczek, just to alphabetically list a few who did. Today, we finally witness the dams breaking and ever more people dare to 'come out': Gravity is describable as the thermodynamics of some microscopic stuff. Relativity is the expected low energy symmetry that emerges from a wide range of condensed state systems, be they crystalline, amorphous, or liquid. Emergent relativity is not a surprising new development but was suspected by those in the know all along. GR orthodoxy held science and philosophy back; its aim of defending good science backfired in multiple ways.

---

[a] "Dynamic" means here interactive (forces rather than plain kinematics) rather than the ability to change, because space-time includes time and there may be no further time dimension to allow change.

Thinking in terms of ether models helps understanding relativity. Those who refuse to consider ethers even as mere didactic tools often fail to understand the weakness of their arguments. If people who are expected to know better in light of their established position nevertheless uphold obviously false arguments, perceptions of establishment conspiracies against the truth are confirmed. Also in order to work against such perceptions, we will here argue for relativity/relationism/abstractness in a unique and perhaps counterintuitive way, namely by emphasizing the relevance of the absolute/ether/space-substance point of view.

In order to reach many educators, we avoid some misleading terminology. Many philosophers regard Newton's space-time as "pseudo-substantial", and this comes usually under the heading of "substantivalism"[3], although Newton's space is clearly an empty box compared to Descartes' space substance. Substantivalism is an "ontic commitment" to the points of space-time and thus naturally fooled in any emergent space-time scenario. Substantivalism and similar are not what we will casually call "substance-views" for lack of still unoccupied terms. The latter includes a variety of concepts ranging from ether fluids up to emergent relativity in modern particle physics. Perceived space-time regions do *not* coincide with regions of particles of an underlying hypothetical substance, *although* substance-like 'stuff' is involved. Asking whether such substance is conserved or even where such space is located itself is not philosophically naïve.

We equally refuse some in the philosophical literature popular concepts. So called hyperplane foliations are misleading and inferior to light-cone descriptions[4] in philosophical debates. A display of great sophistication comes with discussing indeterminism due to the "hole argument[5]", but such are complexities concerning aspects that likely reside outside of the domain of applicability of GR. The thermodynamic character of GR is strongly indicative of an emergent and perhaps even entropic gravity scenario, which would mean that not only curvature singularities, but also worm holes, closed time like loops, and perhaps even the inside of black holes are all merely mathematical peculiarities outside of the domain of applicability of GR, no matter their beauty and consistency.

There are plenty of excellent references[6] if a historical account on closely related topics is desired. The relevance of this work rests in its contributing to the substantivalism versus relationism, absolutism versus relativism debates by supplying an up to date support of relationism and relativism that is enlightened by emergent relativity. We critically summarize well known positions and present some still novel expositions, for example improved arguments against cosmologically preferred reference frames (Section 3.1), the "expansion-paradox" (Section 3.3), and the relational reduction of space-time via the 'non-existence' of light (Section 3.4). Especially Section 4 clearly acknowledges ether-type emergent relativity in modern physics and their didactic value. This includes lay-person accessible explanations of time dilation, causality preserving superluminal propagation, impossibility of time travel, and quantum non-locality not violating Einstein locality.

In the light of modern physics, relation(al)ism is not supported by physics but by philosophy; it is incorrect and misleading to base one's view on contemporary or perhaps even the best possible future physics. No space-substance can be the ultimate foundation. Abstract relationism prevails; alternatives do not make sense fundamentally, and this is independent of the future success of a background independent theory. Note that as much as this argues against space-substances, it allows them. This is a necessary confusion and reflects the important role of so called dualities in modern physics. Dualities played an important role in rendering emergent gravity acceptable. Conceptual duality, i.e. the existence of very different descriptions that seem mutually exclusive but are actually equivalent, are a necessary confusion we need to embrace.

## 2   The wide spectrum from abstract to substance views

GR explains how space-time behaves on large and medium length scales, but it is silent about any underlying microscopic nature. Theories like GR and thermodynamics derive their beauty and strength from being grounded in self-consistency that is independent of the nature or reality of any underlying microscopic physics. GR regards space-time as no more than the relational description that allows space-time events involving energy densities to be consistently arranged relative to one another. This concept of "space-time is pure arrangement" (*abstract-view*) discourages the point of view that space may be successfully described by the assumption of it being "something", some substance whose properties and dependence on (abstract relational) time give rise to it being describable by GR (*substance-view*).

There are alternatives to GR like MOND[7] for instance, which mainly substitutes for dark matter. "Einstein-Aethers"[8] introduce terms to field equations or actions that have effects on small length scales or low accelerations, say in order to modify galaxy rotation rates while preserving Mercury's perihelion shift. Modern ethers like quintessence[9,10], scalar-tensor theory[11,12], dark fluid[13], or Chameleon scalar field[14,15] are motivated by the observations of accelerated cosmic expansion. TeVeS[16] was introduced considering everything from early nucleon synthesis down to the whimsy Pioneer anomaly[b]. Many of these proposals motivate advanced ether-drift experiments[17], which could already justify their classification as substance-views, but partially due to the stigma involved, these proposals have been promoted as abstract and in the tradition of orthodox GR.

It is an open question whether the space-time we perceive is fundamental or maybe a condensed vacuum state[18], a three or more dimensional membrane in a perhaps non-relativistic bulk space-time, a particular layer in a tower of many strata, or even just an illusion computed by a two dimensional sheet. The most novel and promising proposals have relativistic space-time emerge orthogonal to a holographic screen[19,20], which could be a real membrane after all (if it is not enunciated in terms of so called null-surfaces[21]).

Scientists do at times consider implied space-substances but few focus from the outset on such concrete physical things. Especially 20th century physics has been strongly advanced with *concrete* models, by way of considering the *measurement tools*

---

[b] Radio signal data reveal the velocity and distance of spacecraft. After all known forces are taken into consideration, it appears that a very small sunward acceleration of *a = (0.874 ± 0.133) nm/s$^2$* remains for

(operational arguments involving light rays, Heisenberg microscope, instrumentalism), and via principles gleaned from hands-on-physical situations like free falling lifts (weak equivalence), interacting wave packets (uncertainty), and noisy transmission channels (entropy law). This justifies deceptively naïve questions like "If space *were* a substance that has to itself expand or multiply during expansion, what would that imply?"[22] This benefits model-building and is very important. Nevertheless, we are committed to an abstract point of view on the nature of *fundamental* space-time. This lends support to GR, because GR holds on very general assumptions. GR holds as one of possibly several dual descriptions of our perceived space-time and it may hold for a more fundamental space-time, even if our perceived one happens to only emerge inside something comparatively absolute.

## 3 Evaluating arguments for abstract relational space-time

The listing here starts with the generally better known ones (Section 3.1) and strengthens them at times, goes on to consider thermodynamics (Section 3.2), then adds arguments that focus on time and its unification with space (Section 3.3 and 3.4), and finishes with the intuitiveness (Section 3.5) of space-substance models. The items partially overlap and could be put in a different order, e.g. one could employ Occam's razor (G6) to cut an infinite regress short (G5).

---

both Pioneer probes. The product of light velocity $c$ and Hubble constant $H$ is close to $a$. In spite of hyped reports every other year, the Pioneer anomaly has not yet been explained away.

### 3.1 Generally well known arguments

<u>G1) Operational justification:</u> A very important fact is that an *operationally* justified *axiomatic* basis yields Riemann geometry as the observed geometry in the philosophically correct sense of "a-priori"; a term physics often misappropriates. Many authors have dealt with this thoroughly and there is little disagreement relevant here. Considering how we must measure leads to Riemann geometry. This suggests that GR describes fundamental space-time. However, it cannot prove that we experience a physically fundamental space-time, because we must measure with those objects and light rays that emerge along with our perceived space-time (which is fundamental at most in the sense of Kantian phenomenology).

<u>G2) Historical success:</u> Around the time that special relativity (SR) was developed, the famous ether drift experiments[23,24,25] showed that the then prominent ether models were either wrong or hidden from any at the time conceivable experiment. SR's successes and the whole notion of relativity dissolved the substance of space. GR drove this point home further. Event horizon singularities around black holes were resolved to be inconveniently chosen coordinates. This leaves the same flavor behind as a proper explanation of SR's twin-paradox, namely that absolute notions will get you into trouble! Everybody learning GR goes through this struggle, which may be why many fall in love with abstract, orthodox GR, especially if they follow a historical approach. The historical approach focuses on the weakest arguments. The Michelson Morley ether-drift experiments do certainly not prove fundamental relativity.

<u>G3) Significance of cosmic background:</u> GR's localization (gauging) of SR's global Lorentz symmetry is not a symmetry breaking, but GR breaks the symmetry via its cosmology. GR implies (!) a cosmology and its cosmology implies an average background of galaxies and a cosmic microwave background (CMB). This provides access to a unique reference frame in which the background appears isotropic. The universal reference frame is of greater significance than any hypothetical spaceship, and it inspired substance-views via rubber sheet and raisin dough models that illustrate the cosmic expansion. However, GR cosmology never conflicts with the relativistic paradigm (coordinate covariance), which may mean that any imagined ether stays hidden from our observation not only practically, but *on principle*. Still, even on principle unobservable things may still exist meta*physic*ally.

There is no time before classical GR's big-bang. The beginning of the universe is *not* the big-bang[c], but even in case of inflation, there is no classical time to speak of before the beginning of our particular universe. The term "spontaneous" is thus somewhat questionable. Nevertheless, spontaneous symmetry breaking is involved and the analogy with an upright standing stick that falls over applies: it has to fall into some random direction. According to the quantum mechanical (QM) relative state description[26], it falls in all directions. No direction is special then, and this robs the CMB frame of its significance.

G3 touches on several difficult issues, which contributes to its weakness as one cannot explain fully without GR *and* QM. GR has certain cosmological solutions that demand the whole of space-time be given as one object with some consistent topology. Consider two pulses of light being emitted into diametrically opposite directions. In case of some closed (by curvature or topology) universes, the pulses cross each other's paths at the antipode to the event of emission; hence there is yet another preferred cosmological reference frame. Such a cosmic frame exists also without a closed topology: If one wanted to distribute all matter in a universe so that it is equally distributed (in terms of positions and of velocities of all particles inside a Milne model for example) relative to any arbitrary inertial frame S($v$) having velocity $v$ along some $x$-direction say, there would only be one partial solution to it: have all energy in form of light traveling with light velocity; half of it travels to the left and half to the right. Moreover, because of Doppler shifts, the amount should be either zero or infinity, otherwise it cannot be the same for all reference frames. There is no way to have a big-bang start completely Lorentz symmetric and then somehow break symmetry. An upright pen on a horizontal surface has an equal probability to fall in any direction. If Minkowski space is established already however, one cannot put a Lorentz symmetric probability distribution into a future light cone in order to let the CMB frame be spontaneously chosen. Moreover, if one could do so, closed topology and matter content would not necessarily both induce *the same* reference frame. A coincidence of these frames would strongly suggest the preferred frame to be absolute and due to some undiscovered physics.

---

[c] The Hubble expansion or "flow" is extrapolated backwards in time until classically a singularity (zero-size) would have been present. This is the correct big bang and it has been identified with the reheating

G4) Beauty and power of manifest symmetries: Symmetries are power tools in physics, e.g. when applying conservation laws rather than integrating infinitesimal changes over time. Symmetries are the origin of seemingly divine coincidences in descriptions that only realize these symmetries implicitly. Nature appears to conspire against a perpetual motion machine, always throwing in some effect that yet again makes a novel design

---

period after cosmic inflation. Hubble flow is the recession velocity $v = D H$ at any distance $D$ away from unworkable. Theories that make the responsible symmetries explicitly manifest are power- and beautiful for the same reason: their symmetries. The translation of complicated dynamics (forces) into mere kinematics is extremely elegant, like when explaining Lorentz contraction or cosmic expansion (see S2 below). Summarizing all of gravity physics as localized (gauged) group symmetry is extremely beautiful. However, it is no guaranty for ultimate fundamentality in physics. The symmetry of quaternion division algebras and octonion non-associative algebras for example has captivated for centuries, but no fundamental role in physics has been found, and not for lack of trying.

G5) Avoiding infinite regress: If space is ether, where is that ether? If in another space, where and what is that other space? Substance-views favor considering the substance's perhaps higher dimensional embedding, but an ultimately fundamental theory must by definition explain everything without reference to another, more fundamental one. A finite series of spaces in spaces requires an abstract relational foundation for the lowest stratum, which floats 'in the air'. An infinite tower of space-substances as well as a weird circular construction would be itself not a substance-view either, but highly abstract. G5 is the strongest argument for fundamental abstractness. However, it should not be misapplied. Insisting on GR being fundamental cuts an infinite regress short. This is attractive, but possibly cuts one or more steps too short. A higher dimensional embedding was only early on perceived as a nuisance rather than an opportunity. Nowadays, many favor to have as many dimensions in the fundamental space-time as needed to embed all standard model symmetries. The standard model "U(1) x SU(2) x SU(3) x P(3,1)" would ask for at least 1+2+3+3 = 9 space and one time dimensions, which points to 10-

---

the observer and for a given Hubble constant $H$, which depends on cosmic time.

dimensional string theory. Two time dimensions being necessary[27,28] to embed all symmetries together may hint at rather immediate abstractness, because such cannot emerge from a substance living through one time. But again, we should be careful. Prematurely abstract views fail, like for instance the early bootstrap models of the strong force. We may still discover strata that invite to ask what things they consist of (atoms, nucleons, quarks, strings, …) before an abstract relational fundament is reached. Sciences generally (e.g. physics, biology/sociology) suggest emergent phenomena that somewhat decouple from the lower layers they supervene on to occur every few orders of magnitude. There may be much to be found still in the so called "desert" between the nowadays resolved size scales and the by many orders of magnitude smaller Planck scale[d]. Heisenberg uncertainty and the fact that (for high resolution needed) large energy hides itself behind event horizons do not conclusively prove that even the Planck level is the lowest stratum. QM uncertainty can be modeled as due to an underlying medium having a temperature proportional to the Heisenberg constant[29]. Quantization can also be emergent[30,31] or of topological nature (e.g. string winding numbers). Only entanglement in Einstein-Podolsky-Rosen[32] (EPR) situations proves that QM cannot be emergent from a classical foundation like GR can, but that does still not prove that GR is not emergent.

G6) Occam's razor: If different views are equally consistent with all observations, an application of Occam's (Ockham's) razor should be considered: given two equally powerful theories, the more parsimonious one with less assumptions should be favored. If preferred frames and other substance-like aspects of space-time are fundamentally hidden, they could be unnecessary assumptions. Occam's razor is a popular argument and relates to for example Leibniz equivalence, the parsimony of identifying indiscernible states. Many would think it an oversight not to list it, but it is perhaps the worst argument given the ever improving ability to discern previously hidden or unnecessary aspects. Occam's razor must be sparingly applied in order to avoid cutting important aspects, like cosmological time, that further progress may require again. Einstein himself shaved off the cosmological constant $\Lambda$, but we had to reintroduce it. Hidden variables are cut out of QM not because they are hidden or unnecessary, but because they have been shown to not exist consistently! Hidden variables that can exist may point to improved theories, so modern physics keeps its tools as rich as possible[e].

---

[d] However, several proposals like TeV-gravity argue that the Planck size is actually much larger.

### 3.2 Arguments considering thermodynamics

GR has so much in common with thermodynamics (TD) that it basically may be nothing but the TD of space-time[2], which would for example explain why so many proposals (Carlip lists eight[33] different ones) for the nature of black holes (BH) all produce the same dependence between the entropy and the area of the BH. The fact that joining BH always increases the total area $A$ may be no more than the second law of TD, which states that the entropy $S$ in a closed system cannot decrease. GR and TD are each grounded in self-consistency of a few, very general assumptions, independent of the nature and reality of any underlying microscopic layer.

T1) <u>The TD analogy cannot argue against GR:</u> The analogy suggests a microcosm from which GR emerges. However, the statistical mechanics of any microcosm can only give rise to TD as we know it. Thus, the analogy suggests that such can only ever give rise to orthodox GR. Microscopic details only modify TD in nanotechnology (finite system TD). Hence, the analogy rejects drawing on ether theories in order to argue against GR. Instead of supporting space-substance based GR doubters, the analogy even discourages corrections on medium length scales that could influence galaxy rotation rates or account for the Pioneer anomaly. Still, space-substance toy models are useful for considering those corrections to GR that are expected near the Planck length. Thus, T1 supports the general shape of GR, but it does not support fundamental abstractness.

---

[e] This is similar to leaving all terms allowed by symmetry inside a Lagrangian, because statistical

T2) <u>The TD analogy cannot imply substance as a fundamental level:</u> The TD-analogy encourages substance-descriptions as intermediate levels on the way to an abstract resolution. TD emerges from the statistical mechanics of the atomic micro level, which is in a sense a very 'mechanical' substance-view, but it depends on atoms being 'hard spheres' at least in the sense that they cannot heat up. It would not work if they were also made from 'gooey substance all the way down'. Let us not discuss why classical hard spheres are troublesome fundamentally; we already know atoms do not dissipate energy due to QM. QM is again based on self-consistency, not on some mechanics of an underlying substance. Emergence from hidden variables is strictly impossible because of the nature of quantum entanglement. For QM, considering infinite regress (e.g. randomness of randomness[4]) and suchlike results in the same conclusion, namely that the fundamental level is on principle abstract relational.

---

mechanics will populate the spectrum given enough temperature.

### 3.3 Arguments focusing on time being unified with space

The following distinguishes several aspects that are directly concerned with how time and space seem inseparably fused.

<u>S1) Relativity "unifies" space and time:</u> Newtonian (Galilean) "space-time" is space *in* time, living through time. The unique time direction is never contested by different observers. In Minkowski space-time, singling out a time-direction $t$ is almost as arbitrary as labeling a direction as $z$-axis and then claiming that only $x$-$y$-planes are real while the $z$-axis is merely a convenient method of taking the stacking of $x$-$y$-planes into account. Such unifications of forces or dimensions are similar and related to the argument from beauty and power of symmetries. They do not support the abstraction being fundamental. Instead of some profound unification, the space-time mix-up may actually just indicate that perceived space and time emerge *together* via the same phenomenon, for example a finite propagation velocity within a space-substance. Moreover, many cannot accept Lorentz boosts as 'just rotations in (3+1)D'. One cannot even rotate once around in the $x$-$t$ plane without involving infinite velocities.

<u>S2) Abstract space-time resolves the expansion-paradox:</u> When considering a homogeneous universe, classical expansion *through* space and GR's Friedmann-Robertson-Walker (FRW) description fit together seamlessly. A cloud of Newtonian dust expanding *through* space is in the FRW picture described as the universe expanding. There is no locally observable difference between these descriptions. This further example of conspiracy-like symmetry is paradoxical: in the Newtonian/SR description, the underlying space stays the same, uninvolved stage, while in the GR picture, space expands in the concrete sense that there is more of it than before; the latter is obvious when considering closed or compactified[f] universes. The expansion-paradox[22] is: *Space expands globally although it nowhere expands.* The abstract-view resolves this with help of *space-time* not being space *in* time: The four dimensional whole is one unchanging consistent arrangement, the "block universe". The smaller space in the past is simply a different region of the whole. It did *not* grow into the larger space of today; it is still in the past.

This is a beautiful argument for orthodox GR, because the whole growth-of-space problematic vanishes. This is partially why the concept of abstract *metric* expansion is thought superior to the concept of expansion *of* space, which has been argued to be especially problematic[34,35]. Such arguments aim to support an abstract view and orthodox GR, but they fail especially where they favor expansion *through* space instead. In arguing against most ether theories, one should talk about expansion *of* space, because the growth of a hypothetical space-substance is their biggest difficulty; it violates the conservation of substance (continuity equation). For substance-models, the expansion-paradox is a contradiction, because substance cannot by some magic globally appear without being locally supplied somewhere (or everywhere, but still with a *locally* acting mechanism). Substance must flow in from the sides or locally expand or units multiply, perhaps by "raining down" from a higher dimension. There is also the question about the costs of new material. The abstract-view feels easy about that, because mere emptiness comes for free. The semi-classical quantum description ascribes vacuum-energy, but GR does not

---

[f] Space periodically repeats and thus has no boundary but nevertheless finite volume. In a flat and infinite universe, it is harder to argue that there is more total space later in cosmological time $t_c$, because $t_c$ is determined by observation of the average background, or better the temperature $T$ of the CMB. The cosmological principle states that the background is the same everywhere, changing only with $t_c$. Still, two regions at different temperatures $T$ do not violate the principle but just imply that these regions are at

have the time-translation invariance that would demand global energy conservation. GR does not even have gravitational energy.

The rapid inflation in the early universe and the accelerating expansion of the current universe cannot be accounted for by known space-substance theories. For example, G. Chapline admits that his superfluid condensate ether, a very interesting model for black holes, cannot understand "non-stationary space-times"[36], e.g. metric expansion if it is not of the stationary de Sitter type[37]. However, once space-substance is able to procreate, even expansion *of* space cannot argue against it. The expansion scalar of GR's Raychaudhuri equation is formally a rate of local space reproduction[38], or local Hubble constant. It is therefore consistent with orthodox GR that a space-substance procreates locally with a base rate (relative to cosmological time) that is time dilated according to the extrinsic curvature that the Raychaudhuri scalar depends on. Exponential inflation can be modeled with a slow dividing of fundamental units, namely only once every $10^8$ Planck times[22]. It is difficult to dismiss space-substances in general.

S3) Space-substances cannot describe the inside of Black Holes

GR and QM are incompatible because of the curvature singularities that GR predicts, for example those in black holes (BH). At the singularities, GR breaks down to be a theory of physical processes. Operationally, singularities are strictly unphysical, because only infinite energy can provide enough resolution to observe one. Substance models avoid singularities from the outset. Only few abstract proposals, for instance loop quantum gravity[39] (LQG), can claim the same. However, contrasting the success of space-substances for modeling event horizons (EH)[40], they mostly fail to reproduce the internal

---

different times $t_c$. Only a background that is not isotropic clearly violates the principle. One may conclude[22]

of GR black regions like BH. The vanishing of space-substance again contradicts the permanence (conservation law, continuity equation) of the substance. Self-procreating ethers which via a reversed expansion mechanism contract inside BH have not yet been considered in that community. Therefore, as far as we know, ether models fail to describe the BH interior, but this is a weak argument against them. While cosmic expansion is an observed fact, the internal of BH is *not* observable from the outside. There are plenty of models that give rise to the observable features of astronomical BH while being different from GR only on the inside of the BH, or at least they differ only starting from very close to the EH. Moreover, the holographic conjecture can even accommodate that an observer falls *onto* the EH without becoming aware of it. String theory describes this as follows: The observer consists of strings that open up and attach to the EH, which is dual to a rigid surface that the strings collide with. The open strings then spread out on the EH without the observer feeling any different initially. The strings distribute and equilibrate with the EH all around the BH; they "thermalize" on it. The observer interprets this as the tidal forces that start ripping everything apart as she closes in on the singularity. This modern description is simultaneously consistent with GR yet contrary to its orthodox interpretation, which holds that the EH is nothing but empty space.

S4) GR does not imply a unique history: Some ether proponents reject orthodox relativity because it seems to implied fatalism. Block universe presentism does not imply determinism beyond the gravitational sector, which includes masses and gravity forces but not for example the strong force in atomic nuclei. Releasing a compressed spring between the halves of a spinning sphere will push the halves apart and trigger gravity

---

nevertheless that also infinite universes grow their total volume over cosmological time.

waves that could not have arisen from the stable mass distribution of a sphere. Philosophically it is a grave mistake to take QM as the *basis for* (instead of asking how it may be *derived from*) the plurality of possibilities. Nevertheless, it is convenient that one can nowadays point towards a well established, mathematically formalized theory: QM and SR are consistently combined in relativistic quantum field theory (RQFT) and something similar will be achieved with GR. Many worlds interpretations[41] (MWI) can be loosely thought of as in a sense including all possibilities of block universes on an equal footing. QM thereby removes the fatalism and this is independent of whether GR is to be quantized into gravitons or not.

S5) The abstract-view promises a *relational reduction of time.* Some refuse to demote time, because it seems that only "actual" dynamics can develop the future according to physical laws so that the 4D picture "ends up" consistently constrained. Widely accepted terminology misleads that way. One of the most fundamental principles in physics concerns the 'variation' of the action-integral that 'settles' on a stationary solution. This seems to imply a flow of time or some arena where a sort of Darwinian selection for the stationary solution takes place. Again, we hold the argument involving regress without definite termination to be the strongest, and "flow of time" regresses to a meta-time that allows time to flow.

The key to an acceptance of fundamentally abstract space-time is not that time is just as plain as space, but the realization that QM negates plain local realistic interpretations of space[42]. With much poetic license: one may as well accept pseudo Riemannian space-time like one commonly accepts space, because the latter is equally only 'in our heads'. The QM modification of direct realism[4] though is neither yet widely accepted nor pedagogical. Very interesting and accessible however is that SR almost completes the relational reduction of perceived space-time even if there should be other, physically more fundamental space-times. As just given away, this does not prove fundamental rather than emergent relativity either, but it is an interesting abstract viewpoint that deserves to be known widely especially in case relativity turns out to be fundamental after all. Therefore, it is presented as the following, separate sub-section.

### 3.4 Special relativity as a role model for relational reduction

Considering kinematics, SR basically amounts to rules for drawing light paths and hence spatial and time like directions into a space-time diagram. One can understand all the philosophically interesting issues in SR, for example the twin paradox, by drawing Minkowski diagrams without calculating. Nothing surprises about persons having aged differently after having traversed different paths through space-time; one should be surprised if they did not. Minkowski diagrams teach the geometrical nature[g] of Lorentz contractions and thus how SR in a sense gets rid of space-time.

If we want to attain the point of view of the light, namely see the world from the light's own rest frame, we will find that the more we accelerate to travel along with the light, the shorter the travel time becomes due to time dilation and the travel distance's Lorentz-contraction. The light's energy red-shifts until it is undetectable. Traveling with light velocity, light has itself neither time to exist nor is there space for it in between emission

---

[g] Because of the velocity of light, the other parts of a rocket ship never know whether the rocket motor is currently (relative to them) on or off. This literally stretches a pulled ship, which cannot directly lead to the for example *shorter* length, but such *mechanical* aspects are necessary for a finite length to attain a different relative length at all. Inside an Einstein-ether, this leads to an absolute contraction. In a Minkowski diagram, the spatial length of the ship is a cross-section through its world path. The size of a

and reception. QM supports this "non existence" of light: source and receptor exchange interaction quanta (photons). The overall action *A* of systems or processes is a measure for how much own identity independent from observation by the environment they have. Quanta are completely exhausted by a single interaction quantum d$A = h$ (Heisenberg constant). They are only the interaction, the observation, but nothing by themselves. Consider also that "no event takes place in the source itself as a precursor to the click in the counter"[43]. However, it is remarkable already without the quantum aspects. All that happens in this description is that two objects, emitter and receiver, interact without there being time or space directly involved. Nevertheless, if one takes all these interactions between objects and observers together, then from a different point of view, the light may have traveled millions of years over vast stretches of space, and so space-time emerges from space-less, time-less (instantaneous) interactions. Unsurprisingly, many assume SR to be close to *the* fundamental description rather than an emergent deception.

SR would relationally reduce also the ageing of objects in between interactions if they were fundamentally made out of light, but assuming suchlike equals getting carried away. However, one learns two aspects. 1) SR is a role model for relational reductions, for turning apparently concrete things abstract, because it does so via good operational procedure: The measured aspect is ideally not implicitly involved in the way one measures, or worse, part of the measuring device. SR measures space-time distances with the best standard possible on principle, namely one that cannot change its length, because

---

cross-section depends on the angle with which it cuts through the history. The angle depends on the relative velocity of the observer.

it has zero length[h]. 2) In a satisfactory reduction, space-time is expected to be replaced by a web-like relational structure that rips it completely apart from its ordinarily assumed order, much like the spatial order turns 'inside-out' if one tries to imagine that even the farthest star is right on top of our retinas as just discussed. Links between events represent plainly the fact of an interaction, but neither length nor direction. The links in the spin-network[44] of LQG represent areas between the volumes at the nodes and do not get rid of space as much as the SR description here suggests.

### 3.5    Contra the didactic intuitiveness of space-substances

Before advertising the great didactic value of intuitive space-substances in Section 4, let us be aware of their shortcomings particularly in that regard.

I1) Space-substances are misleading: In eternal chaotic inflation, an expanding universe does not squeeze the already present pocket universes outside of it. Infinite space in any one pocket universe can be accommodated by bending it into the time direction. Space-substance models encourage people to plainly refuse accelerated metric expansion on grounds of that expanding substances push towards the outside and squeeze anything inside. Metric expansion can also be interpreted as the significant scale (Planck length) becoming smaller. Substances whose fundamental units shrink or split at a certain rate as given by a local Hubble constant are also concrete physical. However, here we have already an intuitiveness that helps the intelligent model builder while being beyond the comprehension of a wider lay public.

---

[h] Orthodox SR insists on that proper lengths are not some misleading sophistication while the "real" lengths and times are actually different. To insist on a light path's finite length implies a special reference frame. The proper distance along a light path is zero. Not accepting this 'non-existence' of light asks for ether!

I2) Intuitiveness does not imply fundamentality: Physics (P) is the basis underlying the everyday mesoscopic world and corresponding models (M) useful in it: P → M. If in P, like for example in GR, we spot structures (e.g. Einstein-ether) that behave similarly to something in M (substance), a common misconception is that this implies "*M concepts fundamentally underlying P, i.e. M → P. A different universe's $M_j$ leads to $P_j$ instead. $M_i$ (say particles) is helpful in dealing with $P_i$ (say QM) because $P_i$ talks about the properties of $M_i$.*" $M_i$ helps understanding simply because we are acquainted with it. Acquaintance does not stem from the fact that brains and thought structures (memes) evolve in a world that fundamentally consists of $M_i$. Any $P_i$ gives rise to precisely the world in which concepts $M_i$ become familiar, no matter the fundament. In a different universe, $M_j$-structures find use in talking about $P_j$, even if the latter is conceivably based on what would be best described as $M_i$, for example in case $P_j$ is actually based on our $M_i$, say if we run a $P_j$ world in our computers. This is not hairsplitting. The stated misconception is rejected because it claims "*M concepts fundamentally underlying P*". One cannot reject "*something like M may underlie P*" out of hand.

### 3.6 Summary

A simple ranking cannot give justice to the many arguments (G1-6, T1-2, S1-5, Section 3.4, and I1-2) that can be and at times have been directed against space-substances. For example, T1 repels pseudoscientific attacks against GR, but it does not support fundamental abstractness. T2 does but cannot convince those who distrust QM. Almost all arguments, especially the often and widely employed ones, are weak. Only one argument is decisive, because it stands on philosophy instead of resting on partial knowledge of the physical world. G5 considers infinite regress and leaves us only one choice: fundamental space-time is abstract relational. However, it is important to keep in mind that the support of fundamental abstractness cannot be based on cutting the infinite regress short prematurely. Our space-time may not be the fundamental one. To argue G5 correctly, one needs to point out that even allowing a partial, circular, or infinite regress will result in a theoretical construct that is an abstract model far removed from what proponents of intuitive space-substances desire.

## 4 The value of space-substances for physics and philosophy

SR and causality together do *not* preclude all information carrying signals with superluminal speed[45]. Einstein-ether descriptions of SR can teach this easily (Section 4.1). Hardened 'skeptics' refuse employing ethers didactically but also do not spend time considering issues properly within orthodox SR, where it is much more difficult. It confirms perceptions of establishment conspiracy when obviously incorrect arguments defend an orthodox interpretation. This section aims to clearly acknowledge the value of even naïve substance-models while nevertheless staying firmly for the validity of experimentally confirmed GR and for abstract relational fundamental space-time in general.

With the advent of stringy universe-on-a-membrane models[46,47,48], what is in our sense substance-views has entered the main-stream. String theory gravitons interact and thereby 'give rise' to the force of gravity[i]. GR's geodesics through curved space-time had abolished gravitational forces. Rest mass in the standard model, that includes even pure inertia against gravitational acceleration, is described as a permanently ongoing interaction with a Higgs background. Inflaton field dynamics inflates the early universe. Thus, the *dynamics leading to kinematics* concept is deeply rooted in modern physics, so deeply that Richard Feynman abandoned his catchiest motto: "The vacuum is empty"[49].

---

[i] String theory may find a background independent description, but its present language is in stark contrast to the ban against preferred backgrounds in relativity.

Space is some*thing* having properties, as there are gauge field impedances due to for example electro-magnetic (EM) permeabilities, and the fact that space is effectively showing tension and inertia against bending and stretching, namely GR. Space is dragged around in the ergosphere outside of a rotating BH. Space is never empty but teeming with stuff which is not metaphysically 'virtual' or created by observation: Unruh temperature due to acceleration exists without specifically looking for it; BH evaporate via Hawking radiation without conscious observers around to watch. Space might be quite tangible as an actual web of strings or the inside of a super fluid like Helium III[50]. The absence of large scale rotation in the universe follows trivially from super fluids being irrotational. It was already shown[51] in 1945 that a crystal-like Dirac-sea mimics Lorentz contraction and mass-energy increase[j]. That GR could be emergent in a condensed-matter theory has a long history (reviewed elsewhere[52,53,40]). Space-time in GR is similar to stressed matter[54]; there is a close analogy between sound propagation in background hydrodynamic flow and field propagation in curved space-time[55], and so on. The vacuum may indeed be the ground state of a condensed matter system, ordinary matter its excited states.

Relativity emerges because observers are also made from the excitations, are out of pseudo particles of the underlying material. This has been pedagogically well discussed, for example starting with Newtonian fluid dynamics[45]. Substance-views are powerful didactic toy models: Raisin dough illustrates the isotropic Hubble law observed in metric expansion. One can with conveyer belt or fluid pond models quickly recall in front of one's inner eye, i.e. plainly see (visually imagine), how clocks made from excitations of a background undergo time dilation relative to the background *and each other*. As suggested by Landau's dispersion curve below the roton minimum, smoke-ring like vortices in super fluids can carry negative effective mass. Such to velocity anti-parallel aligned momentum has been experimentally confirmed[56]. Collision partners are then pulled rather than pushed away. Therefore, the fact that exchange particles may carry attractive forces can be demonstrated in front of high school pupils long before mentioning QM.

### 4.1 Causality preserving superluminal velocity but no time-travel

Let us exemplify the didactic power of substance models by applying them to two widely discussed issues, namely the light speed limit and time travel. Substance models with emergent relativity enlighten here so effectively that they deserve to be taught widely. If space is like a fluid's surface, the world made up from its low energy waves is special relativistic due to the constant wave velocity $c$. Low energy means wave lengths longer than the liquid's inter particle distance and small amplitudes. High energy phenomena[k] may lead to shock waves (sound), solitary rogue waves superposing non-linearly, wave crests breaking, and fluid splashing[1]. Splashed fluid travels "above" the surface carrying away otherwise unaccounted for energy and momentum. This is the same as the string-theory membrane inspired and well received "particles can be kicked off our 4 dimensional manifold …"[57], and indeed the Large Hadron Collider (LHC) is testing for exactly this type of scenario. Very high energy collisions might let parts of the membrane (also made from strings) come off and travel through the bulk. This suggests an explanation for certain neutrino data and a resolution of the Greisen-Zatsepin-Kuzmin (GZK) cosmic ray paradox similar to but still somewhat more natural than has been proposed before[58,59]. Cosmic rays with energies above ten Joules interact with the CMB to produce pions, thus they cannot travel far, yet some are observed to have five times that energy. If extreme high energy events splash underlying medium, the splashed parts would travel outside of the surface or membrane that makes up the observable universe and thus outside of the CMB. Gravity, as opposed to EM forces, reaches into the bulk next to the membrane and pulls the splashed parts back. While the origin is far away, the secondary sources would be at the re-entry of splashed parts into the observable universe close to the observer, thus circumventing the GZK limit on the travel distance through the universe.

---

[j] A moving Burgers screw dislocation in a crystal contracts to $L' = f L$, where $f^2 = 1-(v/c)^2$ and $c$ is the velocity of transverse sound. The energy is the dislocation's potential energy at rest divided by $f$.
[k] Increasing energy, one will first observe that $c$ becomes dependent on amplitude and wave length. Similar is expected in most QM gravity proposals.

Considering a pond of fluid, splashed drops may reenter the surface after traveling above it with *higher* than the low energy wave speed $c$ observed by observers living inside the surface. Their low energy world is special relativistic[m], yet allows faster than light phenomena that can never violate causality. The splashed substance carries at least the information that a high energy experiment has taken place. Superluminal information carrying phenomena need not violate causality and the substance model shows comprehensively why: the signal travels at most instantaneous relative to the cosmological space-substance (the pond), i.e. it is *tied to one and only one reference frame*! There are inertial systems relative to which the superluminal splashing goes backwards in time according to their time labeling of events, but in none does it splash into their backwards light-cone, which is their causal past. No equations are needed to prove it: We all take baths and know that no wave or splashed drop visits the water's past.

---

[1] Volovic[50] claims that the next lower one of the alternating strata of effective-standard-model-physics and underlying-super-fluid-vacuum in a tower of unknown extend is principally inaccessible from inside any layer. High energy experiments cannot focus pseudo particles so much as to render the underlying fluid locally above its lambda point.

Imagine you watch sentient beings made from a pond's surface waves: the futility of their efforts to invent a time-machine is obvious. A space-substance giving rise to GR in as far as it is confirmed by observations renders the idea of time-travel equally ridiculous: Excitations of a substance cannot visit a previous state of the substance that plainly *does not exist anymore*. Deriving such results as far as they apply to SR from an abstract view point needs many pages, equations, and diagrams[45]. In the case of GR, there is no way to bring such results home within an abstract-view.

---

[m] All objects and observers are made out of waves trapping each other in patterns (pseudo particles). The "light" in the pond-wave-world is the un-trapped waves propagating with *c*. Having no better measure, this light must be used to measure light, thus it always has the same speed *c* also if measured by the patterns that move inside the surface. A pattern moving relative to the liquid's molecules experiences absolute time dilatation: A light-clock is a simple light wave bouncing between mirrors. If the clock moves with close to the speed of light, bouncing light needs much cosmological time to reach the receding front mirror, so it ticks slowly. A Minkowski diagram suffices to establish that moving patterns measure that systems at rest in the pond undergo time dilation relative to them! No observer knows who moves relative to the pond.

Substance models allow for superluminal speeds whenever the speed of what is light inside a particular substance or membrane is slow relative to the (maybe also covariant) bulk space. Our universe perhaps not being a membrane does not touch the main implication, which is that no faster than light travel that is *bound to one reference frame* can violate causality in a specially relativistic universe. Issues that may involve propagations that are *bound to one single frame* include QM non-locality (EPR[32] paradox), "splitting" of worlds due to entanglement in MWI[41], the Scharnhorst[60] effect, and possibly faster than expected[61], perhaps instantaneous[62,63,64] QM tunneling. These do not require much energy, but space-substances do support also infrared effects braking Lorentz invariance[65]. However, never forget that this is not about arguing for ether, but about being able to argue without even a single mathematical formula something as complex and philosophically relevant as that QM non-locality does not violate Einstein-locality[4]. If many world models' branch along hyper surfaces connecting two entangled measurement events (collapse instead of propagating decoherence), every branch has *one* such split surface at its beginning, i.e. only *one* preferred frame in which information may have propagated. These frames (tunnel, split) do not necessarily coincide with a preferred cosmological frame. Also attempts at replacing the Lorentz group by different transformations to provide a kinematical basis for high energy physics that breaks Lorentz invariance (e.g. lower velocity of high frequency light[66,67,68]) usually do *not* argue for fundamental reference frames. For the Scharnhorst effect, the metal plates break Lorentz invariance. If there are two pairs of plates or two tunnel barriers in relative motion, we expect QM arguments similar to Hawking's chronology protection conjecture to show that signals cannot be turned around or reflected so that they end up in the past light cone. However, no experimental evidence disproves instantaneous tunneling or wave-function collapse relative to a cosmological reference frame, as firstly suggested by Hardy[69] and being under experimental investigation[70]. Such is occasionally implied by stating that QM violates the strong equivalence principle[71] and that causality is therefore a global question of topology[72,73]. If tunnel delay time is instantaneous relative to the CMB, one could conceivably test this with 'ether-drift' experiments employing tunnel barriers, as has not been suggested before.